\documentclass{article}

\usepackage{arxiv}

\usepackage[utf8]{inputenc} 
\usepackage[T1]{fontenc}    
\usepackage{hyperref}       
\usepackage{url}            
\usepackage{amsfonts}       
\usepackage{nicefrac}       
\usepackage{microtype}      
\usepackage{lipsum}
\usepackage{graphicx}

\title{Digital Watermarking of Video Streams: Review of the State-Of-The-Art}

\author{
   Romain Artru\\
  École Polytechnique Fédérale de Lausanne\\
  CoSMo Software Pte., Singapore\\
  \texttt{romain.artru@cosmosoftware.io}
  \and
     Ludovic Roux\\
  CoSMo Software Pte., Singapore\\
  \texttt{ludovic.roux@cosmosoftware.io}
  \and
     Touradj Ebrahimi\\
  École Polytechnique Fédérale de Lausanne\\
  Multimedia Signal Processing Group (MMSPG)\\
  \texttt{touradj.ebrahimi@epfl.ch}
}

\begin{document}

\maketitle

\begin{abstract}
Digital Watermarking is an extremely wide aspect of information security, either by its applications, by its properties, or by its designs. In particular, a lot of research has been made about video watermarking and it can make it quite difficult to put into perspective the various schemes possible in order to implement a watermarking process for a given application. This paper presents an in-depth overview of the current video watermarking technologies and how they each respond to certain criteria that may be imposed by the aimed application. The goal being in first place to be able to define the desired equilibrium point between invisibility, robustness and efficiency for an application. Then, given this balance, being able to deduce the best location of the information embedding as well as the method used to embed it. The equilibrium point is to be found using the needed properties of the watermark and by studying the threat model that the scheme will have to face. The location describes whether the extra information should be added to the metadata of the video, to its frames or to specific regions of its frames. Finally, the method to embed the watermark refers to the insertion domain and its coefficients to be altered in order to insert the wanted information.
\end{abstract}


\section{Introduction}
\paragraph{}Digital watermarking is a signal processing technique flowing from paper watermarking, originally being a stamp applied during paper manufacturing by a press as the paper paste was still watery (hence its name). It was first used in Italy in 1282 to prove origin and quality of paper coming from a specific factory located in Fabriano. As the technology discovery went on, paper as well as watermark went from the physical world to the digital one. As the medium changed, new applications (\cite{cox2000}) where found to this process from copyright protection in the multimedia industry to clandestine communications for military use. Those applications will be detailed in Section \ref{applications}. 

\begin{figure}
  \includegraphics[width=\linewidth]{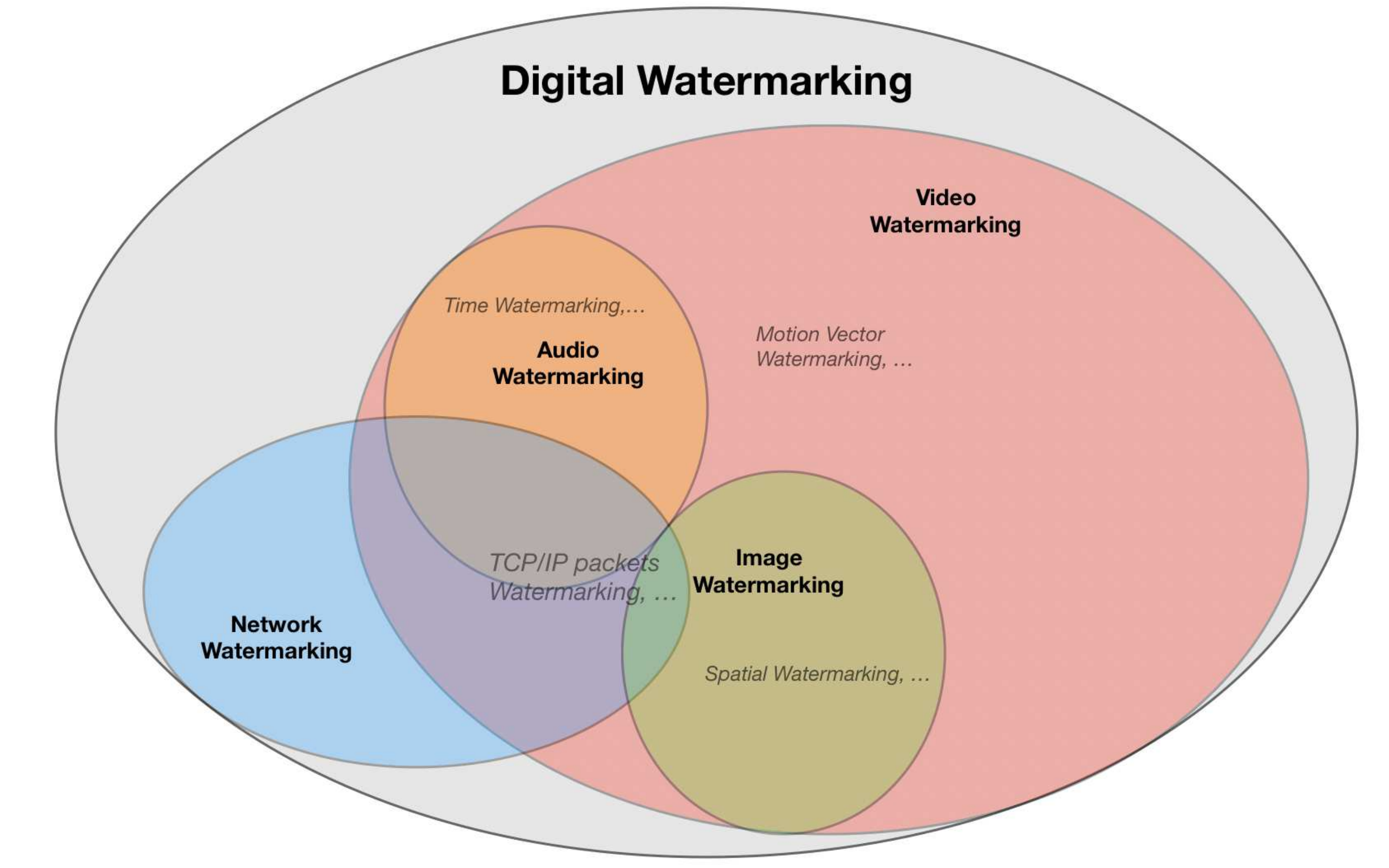}
  \caption{Venn Diagram of main media able to support digital watermarking}
  \label{fig:venn1}
\end{figure}

\paragraph{} We define digital watermarking as embedding extra information (the watermark) into a signal (medium or carrier signal) by using its redundancies. Later on, this signal can be subject to perturbations, malicious (\cite{song2010}) or not (\cite{kuan1985}), and depending on the goal of the watermarking, the status of the watermark should be observable, whether it can still be extracted from the signal or if it has been broken. Concerning the information to be embedded, it can either be related to the medium supporting it, such as inserting a hash of an image in the same image in order to attest its integrity (\cite{wong1998}), or having no relation at all with it, for example to achieve secret communication (\cite{joshi2015}) on a simple radio signal that could be intercepted by anyone. It only depends on what goal is trying to be achieved by the watermarking process. This support medium can be almost anything, the most commonly used being images, but also videos, texts, audio files or softwares are perfectly viable candidates for embedding. The relations between watermarking in those media are displays in Figure \ref{fig:venn1}. 

\paragraph{}This survey however, focuses on watermarks applied to video streams as carrier signals. Video streams can be viewed from many different perspectives that bear various watermarking techniques. We will here consider the following ones:
\begin{itemize}
\item First, one can see it as a sequence of bits. Watermarking being useful mainly when the medium is transmitted between entities of a network, we can also view this sequence as packets of data. Hence, this sequence (or those packets) follows a range of protocols that will integrate some redundancies that can be exploited for watermark embedding (\cite{wang2005packet}). 
\item Another view of videos is as a sequence of frames displayed one after the other at a high rate. Therefore,  any static image watermarking technique can also be applied to videos by applying it independently to the frames of the video. The redundancies used for watermark embedding in image processing are spatial: two pixels in the same region will usually have a higher correlation than two unrelated pixels coming from two completely different parts of an image (\cite{potdar2005}). Those techniques are still perfectly efficient and irreplaceable for some cases of video processing like Real-Time Communications where the following frames are not yet known by the system in charge of embedding the watermark.  
\item As long as the scene change occurs, one could also see a video as one initial static image and a multitude of infinitesimal accumulated changes between the first frame and all the following ones. This view of video sequence induce temporal redundancies that can be used for watermark embedding too (\cite{liu2004}). Indeed, as for the pixels with image watermarking, two consecutive frames will be highly correlated images.
\item Finally, a video usually contains some audio data, which imply that all audio watermarking techniques can also be applied to video watermarking (\cite{dittmann2000}).
\end{itemize}

\paragraph{}This paper discusses applications and techniques currently used in the field of video watermarking. In Section \ref{general}, a general description of watermarking is provided, detailing its properties and applications. In Section \ref{attacks} we present the threat model that video watermarking has to face. In Section \ref{location}, we introduce the main locations where information can be embedded with watermarking. Section \ref{methods} especially describes how to embed the watermark when considering video as visual objects. Finally, Section \ref{technologies} develops how watermarking can be combined with other technologies to provide stronger security.

\section{Watermarking Generalities}

\begin{figure}
  \includegraphics[width=\linewidth]{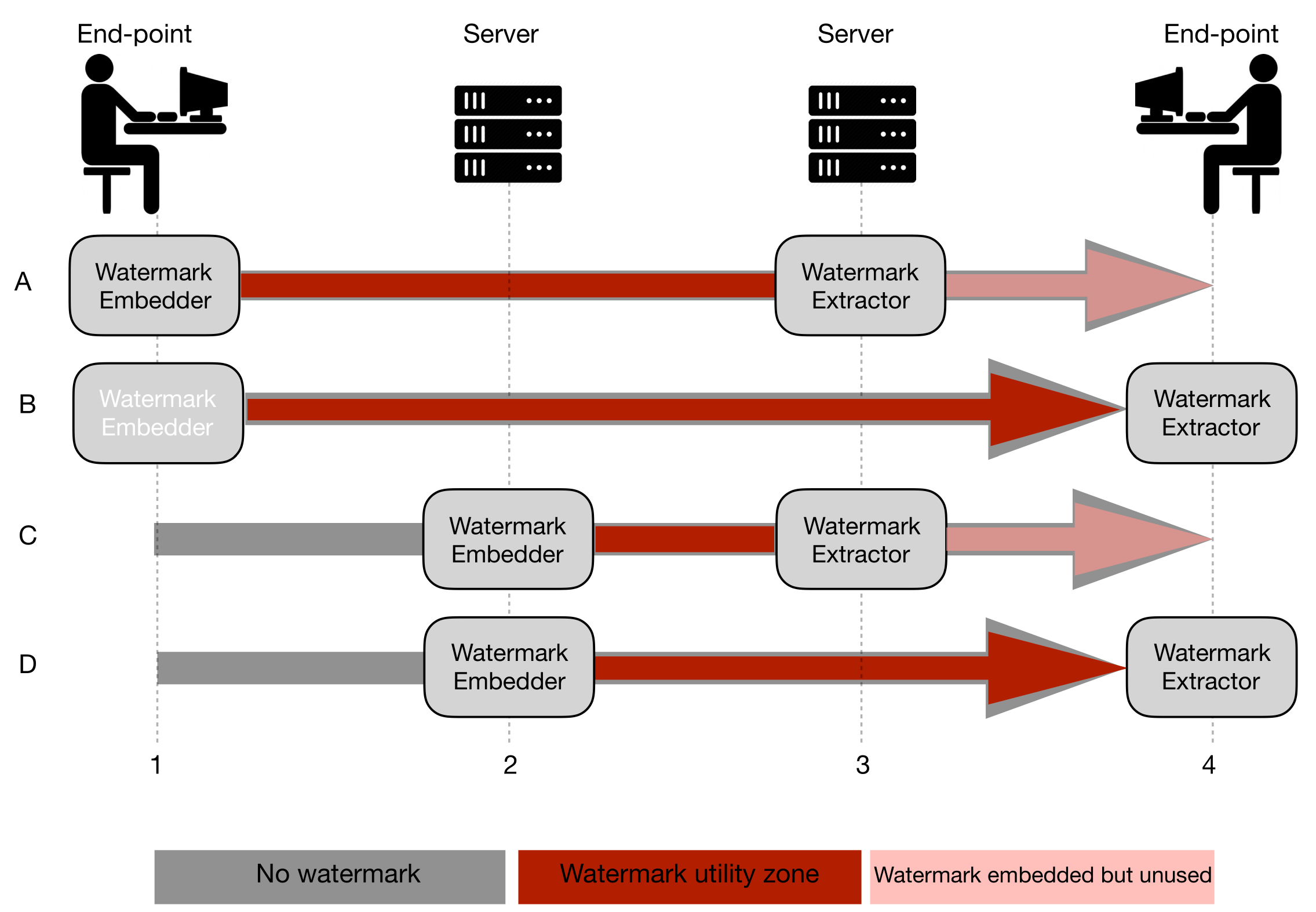}
  \caption{Lifecycles of a watermark}
  \label{fig:lifecycle}
\end{figure}

\paragraph{} We model a video signal as send by a user (end-point) to servers that will then redistribute it to some users. Iacovazzi et al. \cite{iacovazzi2018} extract four different watermark lifecycles from this model. Those are presented in Figure \ref{fig:lifecycle}. We detail Iacovazzi's model by differentiating three possible signal states that can be reached during these watermark lifecycles. Note that the state "Watermark embedded but unused" can be replaced by the state "No watermark" in schemes that allows complete extraction of the watermark and not only detection.

\label{general}
\subsection{Global Scheme}

\begin{figure}
  \includegraphics[width=\linewidth]{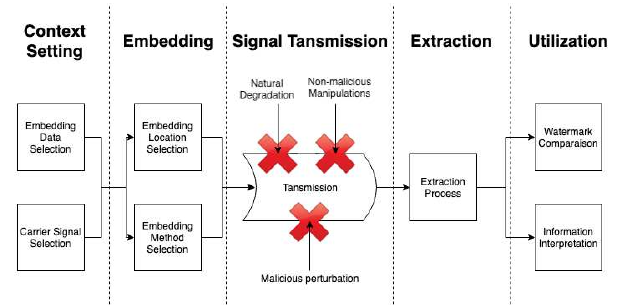}
  \caption{Flow chart describing the general scheme of any watermarking process}
  \label{fig:steps}
\end{figure}

\paragraph{} Since we defined watermarking as embedding extra information into a signal, we model the scheme followed by any watermarking, physical or digital, in five steps as shown in Figure \ref{fig:steps}. The two first phases are mandatory for any watermarking process: the context setting phase as to embed information in a signal one needs to choose which information and what signal, then the embedding of the watermark phase as it is the modification made to the signal that create the watermark. Those two steps are followed by three phases that only happen under certain conditions: for the transmission phase the embedded data needs to be shared, and the two last phases, the watermark extraction and its utilization are executed only when the watermark is exploited. Indeed, one could use the watermarked media by just ignoring the watermark. The first and last phases mostly depend on the application of the watermarking whereas the three other ones define which properties (see Section \ref{properties}) the watermarking will have. In general, the designer of a watermark technique only has access to the embedding and extraction process to ensure the behavior the signal should follow during the transmission phase whereas the person using the technique will have to set the context and treat the extracted data to solve his problem.  

\paragraph{Context Setting}
This step is about observing the application aimed by the watermarking and deducing for it the best way to design the  process. The two most important decisions to take are the selection of the medium that will carry the watermark and what data need to be embedded to fulfill the goal. Even though the medium might seem quite obvious, there can be several possibilities for the same application. Video watermarking is a perfect example to demonstrate this, as we could embed the watermark either in the audio channel or in the visual channel as in \cite{dittmann2000Bis} where both visual and audio channels are watermarked. Concerning the choice of the embedded data, it also directly depends on the application. It could be for example an image, an audio signal, a text or some raw bits. Those two components are the most important inputs of the embedding process that can work as a black box to the user. One might also need to encrypt (or scramble) the watermark before embedding it as \cite{zhaofeng2018} that uses the Arnold transform to do so. This need can come from two facts. First, the user might not want anyone that can extract the watermark to be able to interpret it. Second, to make the watermark invisible as encryption can make it appear as a white noise with zero mean and unit variance. The encryption can be done either by the user of the watermark scheme or by its designer during the embedding process. In the first case the designer simply treats the extra data as raw bits.

\paragraph{Watermark Embedding}
The embedding phase can be seen as a black box, the main inputs being the signal carrier and the data to be embedded (as previously mentioned) and output being the watermarked signal. Depending on the technique used to embed the data, it might also output information needed for the extraction (\cite{gaj2016} output a location map to identify which part of the image was watermarked). We split this process in two parts: the decision of the location where extra data are to be embedded (see Section \ref{location}) and the choice of the method used to add this data to the chosen location (see Section \ref{methods}). Another possible input is a key that can be generated at random or decided by the user to secure the watermark. As the embedding step is the one responsible for the behavior of the signal during transmission and the one that defines how the extracting process will have to work in order to retrieve the correct information, one can consider it as the most important step of the all scheme. 

\paragraph{Signal Transmission}
During the transmission, the signal can be altered unintentionally by natural noise or by users editing it or relaying it. Later on, we will refer to this as natural alteration. The signal can also be modified intentionally by malicious attackers (see Section \ref{attacks}). For those reasons, the scheme should be designed to minimize nature's and malicious attackers effects, and possibly users effects depending on the goal of the watermark.

\paragraph{Information Extraction}
In this step, the watermark is extracted from the carrier signal. However, we consider two different meanings to information extraction: \begin{itemize}
    \item Retrieving a single bit information describing watermark presence or absence in the carrier signal. The extractor will be called binary (\cite{kalker1998}). 
    \item Retrieving the complete watermark extracted from the signal carrier (\cite{yu2002}). We define such extractor as complete.
\end{itemize} 
To achieve information extraction, one can insert additional information apart from the watermarked signal itself such as the logo of the company claiming ownership of the data. This additional information defines the blindness of the watermark and is generally either the original watermark, the original carrier signal before embedding, a key used for embedding or other kind of data such as the location map mentioned in the watermark embedding paragraph (more details in Section \ref{blind}). The extraction, if executed, can be processed at least at three different stages of the watermark lifecycle: at a middle point of the transmission (either by an attacker or by a point relaying the signal), on-the-fly by the receiving end-point, or a posteriori. A posteriori extractions can be processed by two entities: a user trying to remove the watermark (either illegally as an removal attack or legally as part of an access control mechanism), or the owner trying to prove ownership of the product. We observe that in the two first locations the actors are active as their doing might modify the behavior of the receiver's client side, while the last one is passive as it cannot directly influence the way the application will work. 

\paragraph{Extracted Data Utilization}
Finally, the needed information has been retrieved and the last stage of our watermark processing model is starting. If the extractor is binary, then this step is quite straightforward: the user gets a "yes" or "no" answer to one of the those two questions: "Is there a watermark in this signal?" or "Is \textbf{this} watermark present in this signal?". If the extractor is complete, then more possibilities will be available. First, the user can look at the watermark and use it to prove ownership by comparing it to one belonging to the real owner of the signal. This utilization is the same as when using a binary extractor. Second, the user can compare the retrieved watermark to another one in order to determine the transformation that have been operated on the signal. It can be useful to identify tampering of the signal for example. Third, the user can read the watermark as new information on the users that relayed the signal, which is the technique used for network analysis. Finally, the user can use the extracted watermark to retrieve the original carrier signal by simply computing the difference between the two signals. This is referred as "reversible watermarking" and is the preferred approach for Access Control application of watermarking.

\paragraph{}
Another relevant technique to mention as it is a usual watermarking scheme, is dual-watermarking (\cite{mohanty1999, liu2018dual}). It consists in embedding two watermarks using two different methods and in two different locations. This sort of scheme can be particularly stronger than simple watermarking, but usually at the cost of increasing drastically the computational complexity. We will later see examples of such scheme. 

\subsection{Properties and Evaluations}
\label{properties}
\paragraph{} As previously mentioned, watermark system can be viewed as a black box. This black box returns some outputs given a set of inputs. The outputs vary depending on the design of the system that can be described by its properties (\cite{abdullatif2013,panah2016,dutta2017}). Those properties are the how-to of the usage of this black box. The need for one property directly depends on the use case of the watermark. The designer's decision of which properties his/her scheme will fulfill can be difficult, as all of them are linked together, making this decision an optimization problem of the trade-offs between these properties.

\subsubsection{Medium Fidelity (or Distortion)} 
\paragraph{} We define the medium fidelity of the watermark as how noticeable the distortion on the carrier signal after the embedding of the watermark is. In the specific case of image processing, this property is often called invisibility whereas for other signals it is usually called undetectability. There are many measures that can be used to quantify the medium fidelity of a watermark. Most of those measures are detailed in \cite{kutter1999fair}. We here detail the most common ones:
\begin{itemize}
    \item The Hamming Distance: it compares the raw bit streams of the original image and the watermarked image and is defined as the number of bits that differ between them. Also defined the Hamming distortion of sequences in \cite{cover2006}: 
    \[dist(X,\hat{X)} = \sum_i |X_i-\hat{X}_i|\]
    where $X_i$ is the $i^{th}$ bit of the original signal and $\hat{X_i}$ is the $i^{th}$ bit of the watermarked signal. The higher the distance, the more distorted the signal.
    
    \item The Bit-Error-Rate is related to the Hamming Distance. It is given by \[BER=\frac{dist(X,\hat{X})}{len(X)}\] 
    and is the ratio of bits that differs by the total number of bits contained in the signal.
    
    \item The Mean Square Error is also commonly used to describe quality of predictors especially in Machine Learning models as in \cite{das2004}, and is given by the following formula: 
    \[MSE=\frac{1}{n}\sum^{n}_{i=0} (X_i-\hat{X}_i)^2\] 
    The MSE is generally used to assess the quality of a predictor, but can give a first idea of the "error" induced by the watermark in the signal carrier. The higher the MSE, the more distorted the signal.
    
    \item The Peak Signal to Noise Ratio is the most common measure to quantify watermark visibility. It is directly defined by the MSE of the signal: \[PSNR= 10\cdot \log_{10} (\frac{MAX^2}{MSE})\] . In this formula, $MAX$ is the maximum possible value of the signal. The PSNR's unit is the decibel. The lower the PSNR, the more distorted the signal.
    
    \item The Correlation Coefficient will be the last quantification described here. It represents the similarity between the original and the watermarked images and is given by \cite{asuero2006} as \[C(X,\hat{X})=\frac{cov(X,\hat{X})}{\sqrt{var(X)\cdot var(\hat{X})}}\]
    where $cov$ is the covariance between the two signals and $var$ is the variance of the given signal. The higher the correlation is, the more distorted the signal is.
\end{itemize}
\paragraph{} There are a lot of other measures that can be used for distortion measurement, but those are the most commonly used ones in the watermarking field. It is important to note that a watermark user might want a high distortion when the watermark is embedded in order to create an Access Control system for example using reversible watermarking.

\subsubsection{Watermark Fidelity (or Distortion)}
\begin{figure}
  \includegraphics[width=\linewidth]{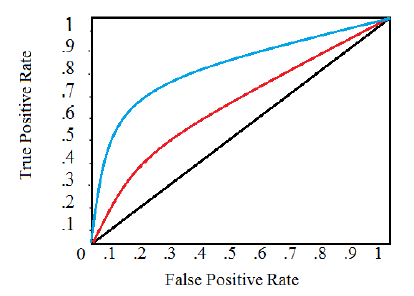}
  \caption{Receiver Operating Characteristics, the diagonal being a completely random decider, and the blue having a better recognizability than the red}
  \label{fig:roc}
\end{figure}

\paragraph{} The name of this property is indeed similar to the previous one, as we use it to describe how the embedded information has been preserved during the transmission phase of the communication process. Its importance depends primarily on whether you need to know the watermarked information for the watermark detection or not. As it deals with distortion, all measurements of medium fidelity also allow to quantify watermark distortion. Another property linked to this one is the recognizability, that our model as well as \cite{kutter1999fair} use to quantify the ability of a binary extractor to output a correct result bit. To measure recognizability, \cite{zafar2017} uses four primitives:  
\begin{itemize}
    \item \textit{True Positives}: number of signal decided as containing a watermark that did contain a watermark.
    \item \textit{True Negatives}: number of signal decided as not containing a watermark that did not contain a watermark.
    \item \textit{False Positives}: number of signal decided as  containing a watermark that did not contain a watermark.
    \item \textit{False Negatives}: number of signal decided as not containing a watermark that did contain a watermark.
\end{itemize} 
\paragraph{} From those primitives, a significant number of values that give information on the recognizability can be computed such as the True Positive/Negative Rate, the Positive/Negative Predictive Rate, the False Positive/Negative Rate, the False Discovery Rate, the False Omission Rate, and the Accuracy. The main way to represent those is using the Receiver Operating Characteristics curve obtained by plotting the TPR (or Sensitivity) against the FPR (or Specificity) as shown in Figure \ref{fig:roc}. The recognizability is observed by considering the area under the curve as explained in \cite{pepe2006}. The closer to the top and left borders the curve is, the more accurate the decider is, the more recognizable the watermarking scheme is.

\subsubsection{Blindness}
\label{blind}
\paragraph{} We call blindness of a watermark the property defined by the prior information needed by the detector to retrieve the wanted data from the carrier channel. \cite{kutter1999fair} differentiate four main possibilities regarding the blindness of a watermark.
\paragraph{Private Watermarking} 
This type requires at least the original signal to be recognized as in \cite{fu2002}. In the case of Network Watermarking (see Section \ref{network}), to determine whether a packet has been embedded by a watermark or not, the original encapsulation and headers of the packets would be needed. This extraction mode can for example compute the difference between the original and potentially watermarked object which should result in an estimation of the watermark (encrypted or not). The watermark itself may or may not be needed depending on the embedding process. So the total inputs of a private extracting process are the watermarked signal, the original signal, potentially the watermark and potentially an encryption key. These watermarking schemes are usually quite robust. Indeed, the more information we have about what we are looking at and for during the extraction phase, the easier it is to determine the watermark's presence.
\paragraph{Semi-private Watermarking}
More often called semi-blind, in such scheme the original signal is not mandatory, but the original watermark is required (\cite{chao2009}). The knowledge given up decrease the robustness of the global system following the same principle as the robustness given by a private watermarking scheme. In \cite{wong2002}, a scheme was designed based on spread spectrum techniques and uses two secret keys to embed the information. Both keys are needed for the detection, but not the original bit stream.
\paragraph{Public Watermarking}
Also called Blind Watermarking, it only uses the received/watermarked signal and an information key to detect and/or extract the watermarked signal. For example, \cite{dutta2017} describe a scheme that uses only a location map to retrieve the embedded information. This kind of technique requires a high level of security to transmit the information key between the embedder and the detector as it is sufficient to extract the data. 
\paragraph{Asymmetric Watermarking}
This last kind of extraction also known as Public Key Watermarking describes a system where anyone with access to the watermarked signal can observe the embedded information, but only one person could have embedded it and no one can remove it. It is by nature computationally more expensive as it usually relies on heavier cryptographic primitives. Schyndel et al. describe a watermarking scheme based on Legendre sequences in\cite{van1999}, their invariance under Fourier Transform allows the author of creating such system. Other schemes can be based on RSA cryptography such as \cite{Lui2018} to achieve asymmetry.
\paragraph{Others}
Some other schemes exist, but they usually define themselves as a combination of some of the four previous watermarkings such as the dual-watermarking method presented in \cite{ahmed2005}, where a first asymmetric watermark is embedded, followed by a second symmetric one. 

\subsubsection{Robustness}
\paragraph{} The robustness property can be defined by how difficult it is to remove the watermark from the embedded signal, whether it is by a malicious attacker, by a non-malicious one, or by natural degradation of the signal. Three kinds of robustness levels are usually distinguished (\cite{abdullatif2013}). They all have different characteristics and can be linked to various specific applications as we will see in Section \ref{applications}. In many of the currently existing techniques, watermarking security is ensured by obscurity, which is completely against Kerckhoff's principle (\cite{Petitcolas2011}). For example watermarks using techniques such as Least-Significant-Bit embedding can be easily detected, extracted and removed if the adversary is aware of how the watermark has been embedded. This is a really important design flaw, therefore fewer and fewer schemes use this kind of insecure methods. Robustness deals more about general signal manipulations as detailed in Section \ref{attacks}. 
\paragraph{Fragile Watermarking}
This level of security is the weakest of all, meaning that the embedded information can be removed very easily. Indeed, almost any manipulation applied on the media would destroy the watermark. The goal is to make sure that absolutely nothing altered the integrity of the signal. In \cite{qin2017}, the author describes a fragile scheme that allows not only to detect the location where the image has been tampered, but also to reconstruct the original image. This is possible using the concept of self-referenced watermark described in the same article: the embedded information is a description of the carrier image.

\paragraph{Semi-fragile Watermarking} 
Related to fragile watermarking, semi-fragile techniques ensure that the signal was not altered in a significant way such as image forgery on top of it or geometric transformations, but will still resist common light signal modifications such as filtering or compression. If one of those light manipulations is applied, the watermark will suffer very few changes whereas if a heavy editing of the signal is executed, the watermark will be destroyed or significantly damaged. This is particularly important for media as the signal is usually compressed and encoded, sometimes with losses in order to be stored and transmitted using less bandwidth. \cite{Kaur2019} proposes an example of such watermarking system. Indeed their results show good robustness against re-compression, noise addition and frame dropping attacks. Moreover, malicious attacks (non-content preserving) can be detected and localized in a video frame.

\paragraph{Robust Watermarking} 
Finally, robust watermarking refers to schemes that embed data as securely as possible so that it is hard to remove the watermark. The main use for this level of robustness is copyright protection as the mark embedded by the owner should never be removed. High robustness can be achieved by various means including embedding at low bit-rate, multiple embeddings of the same information or self-referenced watermarks. An example of such robust watermarking scheme is \cite{parah2016}, which implements a watermarking solution presenting good result against important cropping, rotation, scaling as well as combined attacks such as rotation, cropping and histogram equalization at the same time. 

\paragraph{} To evaluate the robustness, the usual strategy is to apply various kinds of attacks on an embedded signal, extract it and evaluate the difference between the originally embedded watermark with the extracted one. A generic tool has been developed for image watermarking since 1997 called Stirmark \cite{petitcolas1998, petitcolas2000}. It applies a series of random bi-linear geometric distortions to generic images containing a watermark in order to try to damage the embedded data of a given algorithm.

\subsubsection{Capacity}
\paragraph{} Also sometimes called payload, the capacity of a watermark as defined by \cite{abdullatif2013} is the quantity of information that the scheme is able to embed in the carrier signal. It is usually quantified in bits per covert signal unit. If the covert signal is an image or a video, then the capacity is the number of bits that fit per carrier image (or frame). An important trade-off is to be decided between capacity and medium distortion: indeed, the more bits we want to embed, the more visible the distortion induced usually is (at least when using the same embedding technique).

\subsubsection{Time Complexity}
\paragraph{} The time complexity of a watermarking scheme can be divided into two parts: time complexity of the embedding process and time complexity of the extracting process. Their meaning are quite straightforward: the embedding (respectively extracting) complexity is the time that it takes to embed (respectively extract) a signal. When the embedding is executed right before emission (and respectively extraction on reception), time complexity is particularly important as it represents a delay. A common way to quantify the time complexity, especially for the video watermarking, is the Bit Increased Rate (\%) as defined in \cite{dutta2017}: \[BIR=\frac{R_{\hat{X}}-R_X}{R_X}\times 100\] where $R_X$ is the bit-rate of the original stream and $R_{\hat{X}}$ is the bit-rate of the watermarked stream.

\subsection{Applications} 
\label{applications}
\begin{figure}
  \includegraphics[width=\linewidth]{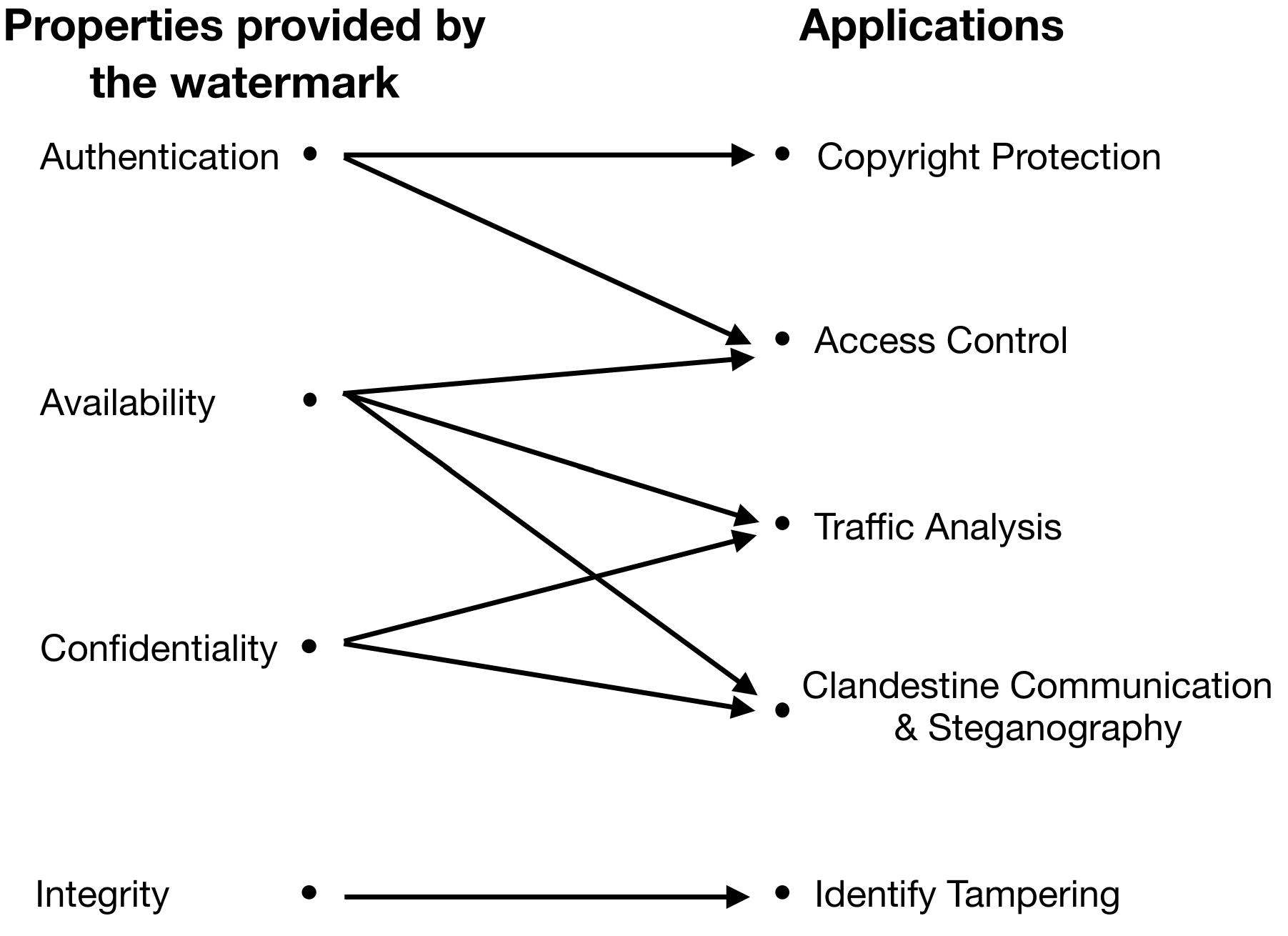}
  \caption{Graph of watermarking applications and the fields they can be applied to}
  \label{fig:app1}
\end{figure}

\paragraph{} Applications of digital watermarking are as broad as the techniques that can be used to implement it. During our research, five major applications stood out as the most encountered ones. These are copyright protection, tampering identification, traffic analysis, clandestine communication and access control. These applications all use digital watermarking to solve some of the four central concepts of information security as defined by \cite{whitman2011}: 
\begin{itemize}
    \item Confidentiality: the information is not disclosed to unauthorized entities.
    \item Integrity: the information is proven to be complete and accurate as the source emitted it.
    \item Authentication or Non-Repudiation: the source that emitted the information prove as well as not deny having sent it.
    \item Availability: the information is available when needed by the authorized entities.
\end{itemize}
\paragraph{} The relationships between the security concepts and the main applications of watermarking are shown in Figure \ref{fig:app1}.

\subsubsection{Copyright Protection}
Copyright protection is the most common application of watermarking. Indeed, as seen in the introduction, proving origin was the main reason why paper watermarks where first invented \cite{hartung1999multimedia}. The basic idea is to embed the information to identify the owner in the product. Then, if a product's origin generates a controversy, the embedded information simply has to be extracted in order to determine who the product belongs to. When the watermark is visible, the extraction process is not necessary as the identifier is directly visible. 

\paragraph{}  The application fields are many, including obviously Art, as the author of a piece, would it be a music \cite{czerwinski2007}, a painting or a photography does not want his creation to be claimed by someone else \cite{lipinski2011}, Geographic Information Systems as geographical data are hard and expensive to collect \cite{voigt2002}. The identifying information embedded can be used in court in case of issue regarding an object's origin under certain conditions as exposed in \cite{Schellekens2014}.

\subsubsection{Tampering Identification}
\paragraph{} To stay in the legal field, one can also use watermarking in order to detect data tampering. For example this can be used to detect forgeries or fake images as they are usually re-assembled parts of images. There are two main ways to identify such tampering on a medium: 
\begin{itemize}
    \item The first one is not specific to watermarking, but is the most common method to ensure integrity: the use of a checksum (\cite{cohen1987cryptographic}). Meaning a string of fixed length that is absolutely specific to a sequence of bits. The slightest modification to the image would change completely the output of the checksum. Linking it to watermarking is quite straightforward: one just has to compute the checksum and embed it robustly into the image. To verify the integrity, the checksum is extracted and compared with the one computed from the original image. If both differs, then the image has been tampered. \cite{sreejith2017} proposes such method especially applied to the medical field where the checksum of most important zone of the image (Region Of Interest) is embedded in the rest of the image. The positive side of the technique is that it only requires a low watermarking capacity as checksums are digest of the global images, the counterpart is that one can only notice that the image was modified, but not where nor how.
    \item The second technique specifically uses the properties of watermarking as it relies on a low robustness of the watermark. If the image is modified, so is the watermark. Using this principle, one can not only observe that the image has indeed been modified, but also where it has been tampered. The original carrier signal can then sometimes be recovered. For example if the watermark was a self-referenced one as in \cite{qin2017}.
\end{itemize}
\paragraph{} As we saw, tampering identification has applications in the medical domain, but also in justice where the integrity of images used in court is essential, in military as intelligence has to certify that, for example, pictures used for strategic decision have not by compromised. Journalism can also use this application as it is also a reporter's job to ensure the veracity of his images. 

\subsubsection{Clandestine Communication}
\paragraph{} Clandestine communication using digital watermarks is known as Steganography. Steganography can be considered as a sub-domain of watermarking, since its goal is to embed a secret message into a covert medium. The goal of steganography is often described by the Prisoner's Problem \cite{simmons1984}. This problem describes a situation in which two prisoners must find a way to communicate secretly together in full view of the warden. The main difference with usual watermarking is the importance of the secrecy of message embedded. Steganography usually does not give importance to the covert medium as long as it safely protects the message, we say that steganography is watermark-oriented. The three most relevant properties of a steganographic system are a high capacity, a high undetectability and a low watermark distortion. From this definition, we define watermarking focused on the medium as carrier signal-oriented which relies on a high robustness, blindness of the extraction and a high medium fidelity.  

\paragraph{} There are many reasons to develop clandestine ways of communicating. In the military and intelligence fields, it is very important to be able to communicate between two points, keeping not only the content of the discussion secret to the enemy, but also the presence and location of the two entities communicating. The same reasons can motivate journalists or whistle-blowers to use clandestine communication in order to avoid censorship control. An example of technology designed specifically to circumvent censorship is \cite{invernizzi2013}, where the Message-In-A-Bottle protocol is defined in order to establish first contact between two entities through photos included in blog posts so that they can then start a secure communication. 

\paragraph{} Another application of watermarking that can be considered as clandestine communication is data exfiltration. A well known cyber-attack model is the Advanced Persistent Threats (APTs) \cite{tankard2011advanced}. The model decompose the attack into several stages, the last one being data exfiltration  that aims at retrieving the information extracted for the target computer network. Indeed, some security mechanisms are usually in place to prevent such exfiltration, using steganography is a solution to bypass those security mechanisms as in \cite{iacovazzi2018}. 

\subsubsection{Traffic Analysis}
\paragraph{} As a goal of images and videos is to relay information, these data are usually transmitted. Transmitting data generates traffic and flows that can be analyzed in order to monitor and enhance control and security over the resulting network. Watermarking is one of the technology that allows such traffic analysis. For example, to break anonymity on network traffic, a unique identifier can be assigned to each entity of the network. These ids being automatically embedded as watermarks into the transmitted packets. This application is a part of the fingerprinting technology \cite{cox2002}. One can also automatically detect previously embedded watermark to monitor broadcast of a commercial or a movie. The implementation of traffic analysis using watermarking gives access to an extremely wide variety of tools such as unusual traffic detection, geographical prediction, network design decision making and more as detailed in \cite{iacovazzi2018}.

\subsubsection{Access Control}
\paragraph{} The last presented application of digital watermarking is Access Control. Part of this application is included in the previous description of clandestine communication, as hiding a signal in a radio transmission, for example, restrain access to this information to those unaware of its presence. Access control can also be guaranteed by the implementation of a software client side which blocks a media if it contains (or not) a certain watermark. More information on this application can be found in \cite{petrovic2013}. The domains where such access control is used include TV broadcasting, access to medical information, or network design. 

\section{Threat Models}
\label{attacks}

\paragraph{} Through its lifecycle, the watermarked signal is subject to various degradations \cite{sklar2001}. Those can either be legitimate (voluntary or not) or malicious (performed by a third party behaving in an unexpected and unwanted manner regarding the signal). We consider that video data at rest on the sending end-point before watermarking are safe as threats in this phase do not rely on security of the watermarking scheme.

\subsection{Legitimate Threats}

\subsubsection{Natural}
\paragraph{} These degradations are due to two causes: 1) being subject to physical laws and 2) being part of a network. Even when data is transmitted by physical link, the signal is never exactly identical on both ends.

\paragraph{Physical causes}
The first kind of natural deteriorations happens because no link is ideal and perfectly isolated: it is subject to external effects that can possibly create three different threats to watermarking schemes:
\begin{itemize}
    \item Additive white Gaussian noise (AWGN): A common effect of transmission is the presence of random noise added to the signal. Those can damage bits carrying watermarked information.
    \item Signal attenuation: Also known as signal damping, the amplitude of an analog signal transmitted tends to see its magnitude decrease which can make the watermark harder and harder to detect, especially when it is using techniques such as LSB (see Section \ref{spatial}).  
    \item Interferences: Especially present in wireless communications, interferences refer to the effect other signals having a bandwidth overlapping with our can have on the transmitted data. It usually manifests itself as a distortion of the transmitted media at the bit level.
\end{itemize}

\paragraph{Network causes}
Being part of a network is not without consequences: in order for the transmission to be successful, the signal usually passes by multiple nodes of the network before reaching its destination. There are three main consequences that can affect the watermark:
\begin{itemize}
    \item Packet Loss: The nodes of a network use a given protocol to forward a packet. For various reasons, packets can be dropped by this protocol \cite{bonaventure2011computer}: whether it is because a buffer was already full when it received a packet, or because it was identified as not fit to be transmitted. Losing a packet can be harmless for certain kinds of watermarking, especially if the protocol used allows packets retransmission, but it can also break a watermark scheme if for example, it relies on timing channels for the embedding. Note that this kind of threat can also be voluntary and malicious if the attacker randomly drop packets or frames in order to destroy a watermark.
    \item (Re-)Compression: An important part of signal transmission is compression, the more the signal can be compressed, the faster data can transmitted \cite{tekalp2015digital}. But this compression can be lossy and discard information considered as irrelevant. However, some watermarking scheme rely on that very same information for embedding.
    \item Transcoding: As all nodes of a network do not have the same configuration, packets might have to change their format at some layers during the transmission in order to be forwarded correctly as explained in \cite{xin2005}. But like for compression, only relevant information will be mapped to the new format. In storage channel-based network watermarking schemes, the locations chosen to carry the watermark have to be decided accordingly.
\end{itemize}

\subsubsection{Voluntary}
\paragraph{} These threats represent the tools available to the user of the signal to make it compliant with the goal he is trying to achieve. There are two main families of such manipulations: 
\begin{itemize}
    \item Geometric manipulations: Also sometimes called RST manipulations, they include Rotation, Scaling and Translation. These operations often used by the signal user have the potential to destroy an embedded watermark and should be taken into account when designing of the scheme.
    \item Signal enhancements: This type of operation consider attempts of the user to improve the quality of the signal (for example an image) often by applying  filters to it. As mentioned before, noise might have been added to the signal during transmission. An enhancement can therefore be to remove that noise, a process also called de-noising. In \cite{fu2017}, the authors develop a technique to remove rain from pictures using deep neural networks using such enhancement manipulations.
\end{itemize}

\subsection{Malicious threats}
\paragraph{} A type of attack is often used as a primitive and derived into many others attacks: watermark estimation. There are two main ways to estimate a watermark:
\begin{itemize}
    \item Chosen signal attack: a known signal is being watermarked by the system of which we want to estimate the watermark. The study of the differences between the two signals allows to determine an estimated watermark. This technique has been implemented for network flow watermarking in \cite{lin2012}. However, it is also possible to design it for media watermarking. A blind version of this attack exists. It relies on removal attacks to get an estimated version of the original medium then used to estimate the watermark. This technique is very approximate and shows very poor results but can be enhanced by having prior knowledge of the signal's statistics.
    \item Multi-signals attack: the watermark is estimated using many different signals embedded with the same watermark and the same parameters as described in \cite{dekel2017}. To counter this attack, a common technique is to slightly modify the watermark embedded at each insertion (for example using a small rotation on the watermark). It is also one variant of a collusion attack that will be mentioned later in this section.  
\end{itemize}

\subsubsection{Embedding (or Protocol) attacks} 
\paragraph{} This kind of attack aims to induce a false watermark detection. The three principal embedding attacks are:
\begin{itemize}
    \item Copy attack: This is an example of an attack directly derived from watermark estimation. The goal is to predict the watermark of a given signal, then to copy this estimation on a non-embedded signal. Such technique is, for example, described in \cite{kutter2000}.
    
    \item Ambiguity attack: The idea of this attack as presented in \cite{hartung1999spread} is to create an ambiguity on the watermark extracted. Indeed, one possibility is to implement a detector that can extract a fake watermark that was not embedded from the same signal, which leads to the definition of non-inversibility. A property meaning that one can not prove through extraction that a watermark is present in a non-watermarked image. More specifications and techniques to achieve non-invertibility are given in \cite{craver1997}. Another possibility is to embed multiple watermarks in the signal, then claim that the original watermark is yours, or that one watermark is not more legitimate than another. Some counter-measures to this threat are presented in \cite{singh2017}.
    
    \item Rewatermarking attack: In the case where a fragile watermark is embedded to ensure integrity of an image, an attacker can estimate the watermark, modify the medium as he/she will, then re-embed the watermark to fool the detector that will believe the medium has not been tampered with given that the watermark is indeed detected \cite{nin2013digital, singh2017}.
\end{itemize}

\subsubsection{Detection attacks}
\paragraph{} This second category of threats focus on enabling a party to detect a watermark embedding even though it was not supposed to. For obvious reasons, this technique is not studied for visible watermarks as anyone can immediately detect such watermarks. A detection attack scheme can also include an embedding phase: this is a widely used technique for desanonymization of network flows. \cite{yu2007} presents an example of such scheme where a first attacker embeds a watermark in the signal close to the sender. Then, an accomplice of him/her will know the origin of the signal if he/she detects the watermark somewhere else in the network.
\begin{itemize}
    \item Correlation-based attack: Many variants of this attack are possible depending on the assumptions made. If the attacker already knows the watermark that is suspected to be embedded, he can simply compute the correlation between the signal (or blocks) and the watermark. If the watermark is indeed embedded, the attacker might observe a peak in the correlation. A form of this attack and a response to it is given in \cite{singh2014}. A variant of this attack scheme for network flows explained in \cite{jia2013} uses Mean-Square Auto-Correlation to detect watermarks embedded on Direct Sequence Spread Spectrum (see Section \ref{DSSS}) in order to break desanonymization scheme as the one described above. 
    \item Timing analysis attack: This type of attack is more specific to network flow watermarking, and especially timing channel-based watermarking (see Section \ref{timing}). It consists of observing the rates of packets and evaluating the timing shape, entropy and regularity of the packets arrival in order to detect patterns in the traffic. Those evaluations and detection methods are described in \cite{archibald2012, ghosal2013}.
    \item Deep Packet Inspection attack: DPI is a technique that examine packets in detail before relaying them. It is generally used as a firewall technique, but one could also use it to observe patterns in transmitted packets to identify watermarks embedded using storage or application-protocol channels (see Section \ref{storage} and \ref{AP}). 
\end{itemize}

\subsubsection{Removal attacks} 
\paragraph{} The most common kind of attack however, aim at destroying the watermark embedded in a signal so that this signal can be used freely without being detected as copyrighted for example. Watermark removal attacks often work as a combination of other attacks: first the watermark is detected, then estimated so that it can finally be removed. Many different schemes have been developed to remove watermarks:
\begin{itemize}
    \item Pathological distortion attack: Being the most basic removal attack, pathological distortion attack simply applies signal processing operations that maintain the fidelity of the embedded medium to remove the watermark. Generally, those distortions can be either linear filtering, noise removal, noise addition, geometric or temporal manipulations.
    \item Collusion attack: This scheme of attack has two main forms, the first one has been explained in the watermark estimation part, the second one is more focused on estimating the medium based on multiple copies of this medium embedded with different watermarks.\cite{wang2018} describes for example how multiple malicious clients receiving the same media embedded with personalized watermark can collaborate to remove this watermark and retrieve the original media.
    \item Oracle attack: The basic idea of this threat relies on the assumption that the attacker has at his/her disposal an oracle that can identify whether a watermark is embedded in the signal or not. A detection attack scheme can be used to implement such oracle for example. Once the attacker has this oracle, he/she iteratively apply small modifications to the signal and passes it through the oracle until the watermark is not recognized anymore. Recently, such techniques have been combined with machine learning to improve their efficiency (\cite{quiring2017,quiring2018}). Again, two variants of this attack exist: one based on sensitivity analysis where a binary oracle detects how the modifications influenced the detection (\cite{zhang2007}), and one based on gradient descent relies on an optimization model using statistics of the detection to converge to the original signal as in \cite{quiring2018bis}. 
    \item Desynchronization attack: Even through this attack is not really about "removing" or "destroying" the watermark, it fits in this category as it consists of creating synchronization errors between the watermark embedder and extractor, preventing the detector to identify correctly the watermark. \cite{wang2011} gives example of such attacks and methods to counter those threats. 
    \item Mosaic attack: In the case where a detector is implemented in a server C between two clients A and B. The mosaic attack's goal is to send a watermarked media from A to B through C without its watermark to be identified by the detector. To do so, A divides the media (perhaps an image) into blocks and transmits those blocks independently to B. As each block will only contain a fraction of the watermark, it will not be detected. This attack scheme is explained in \cite{petitcolas1999}. It is also sometimes categorized as a system attack along with pixel scrambling (\cite{tanha2012}) which randomly switches neighboring pixel values to destroy watermarks. 
\end{itemize}

\subsubsection{Cryptographic attacks}
\paragraph{} This last type of attack aims at breaking the cryptographic system on which the watermarking scheme relies \cite{venugopala2016attacks}. It means acquiring a complete knowledge of the protocol and the inputs used for embedding and extraction. This gives to the attacker total control on the watermarking scheme. He/She can either embed new watermarks, remove existing ones or detect old ones. The simplest method to achieve this is to use brute-force in order to get all the correct parameters such as embedding keys. However, those attacks usually have an extremely high computational cost which makes them impracticable.

\section{Embedding Location Selection}
\label{location}

\paragraph{} Watermarking can be achieved at three different levels: the visual level (including static image and sequenced frames), the sound level or the packet level. Even though all of them represent valid potential watermarking media, their implementations and resulting properties are very different. We will not talk about audio watermarking in videos as these techniques provide a very low capacity without any major benefit. One would almost always rather embed the watermark at the visual or packet level. Moreover, audio is not a mandatory feature of videos. For both visual and packet level embedding, the watermark can be inserted in various locations that will be detailed below.

\subsection{Videos as visual objects and The Human Visual System}
\label{HVS}
\paragraph{} To embed a watermark in a frame, the usual scheme is to break the images into blocks, select which blocks are to be embedded, embed them, then reassemble them all together to re-form the image. This section focuses on the selection of the blocks to embed.

\paragraph{} The Human Visual System (HVS) is a model describing the capacities and limits of the human sensory system (\cite{delaigle2002}). By its study, many researches have defined optimized locations for embedding so that the watermark will affect as little as possible the quality of the image to the human eye. Note that compression algorithms often induce losses in the same regions for similar reasons: after decompression, the human eye is less likely to detect those losses. 

\paragraph{} For the embedding location decision process, the features described below can be combined at will in order to select the degree of imperceptibility wanted. However, each feature taken into account usually increases the computational cost and decreases the embedding capacity. The two extremes being to embed every block with only few data and to embed only blocks corresponding to every criteria of the HVS. 

\subsubsection{Color Sensitivity}
\paragraph{} To represent an image, a common method is to decompose the frame into three planes: red, blue and green. This representation is called the RGB standard (\cite{susstrunk1999}). As described in \cite{vaishnavi2015}, the color sensors of the human eye are divided into those sensitive to red, representing 65\% of them, those sensitive to green, representing 33\% of them and only 2 remaining percents are sensitive to the blue color. For this reason, watermarking the blue plane (or channel) induces a better imperceptibility to the HVS. Many watermarking techniques rely on this fact and select blocks only in the blue plane for embedding (\cite{vaishnavi2015, Nakano2010, tabassum2012}). 

\subsubsection{Luminance Sensitivity}
\paragraph{} Another representation of images called YUV or YCrCb defines a color using three channels: the Y component, being the brightness (or luma) and the U and the V components define the chrominance of the color. \cite{lakshmi2017}, as many other researches, states that the eye is also less sensitive to the noise caused by the embedding if it is located in regions with high brightness. Which explains another popular location for watermark embedding: blocks having a Y component with a high value as the scheme described in \cite{Jadhav2014, gupta2018, lu2013,huang1998}. 

\subsubsection{Texture Sensitivity}
\paragraph{} The most exploited characteristic of the HVS, however, is its sensitivity to texture. Indeed, the eye is less sensitive to changes in very detailed regions. Therefore, it is preferred to embed a watermark in textured and edge regions instead of plain ones. To identify blocks highly textured, each block is transformed into the frequency domain using a transform such as the Discrete Cosine Transform (DCT). In such a domain, blocks having high frequency coefficients will reflect texture regions containing a lot of details (\cite{Kaur2015, Nakano2010}). Another technique to detect highly textured blocks is to count the number of non-zero (or NNZ) frequency coefficients in the block (\cite{Kaur2019,abdi2017, dutta2017}). The largest this number, the more this block represents a detailed area. The human eye will hardly notice changes in such regions. Hence this is a widely used embedding location. 

\subsubsection{Motion Sensitivity}
\paragraph{} Another feature of the HVS useful for watermarking purposes is the motion sensitivity: the eye has difficulties to detect changes occurring in moving blocks \cite{zhang2001}. Moreover, embedding a watermark without considering motion might cause temporal flicker of the image. This feature separate video watermarking from image watermarking. Blocks from a given frame are compared with blocks of previous frames to extract motion information. As described in \cite{mansouri2010}, one can use computations to obtain the Normalized Motion Activity value of a block. The higher the NMA, the more the block is moving. Hence preferred embedding blocks have high NMA values. One can also compute the Motion Vector of a block as in \cite{dutta2017, gaj2016} to estimate the movement of a block as a vector and embed the watermark in blocks having their motion vector's norm higher than a configured threshold.

\subsubsection{Watermarking and encoding}
To introduce this type of watermarking, it is necessary to explain few concepts of video encoding and compression detailed in \cite{agi1996empirical}. The first of them being inter-prediction. The idea is to represent the video not as a sequence of frames, but as a sequence of Groups of Picture (GOP). A GOP can be composed of:
\begin{itemize}
    \item I-frames: Standing for Intra-coded frame, it is the first frame of a GOP and is completely independent of other frames.
    \item P-frames: Standing for Predictive coded frame, it defines itself as differences between the frame it represents and a previous one (either I or P).
    \item B-frames: Standing for Bipredictive coded frame, it uses its relationship with a previous and a future frame to describe itself.
\end{itemize}
\paragraph{} Many watermarking schemes use this concept in their embedding location decision process. Some choose to embed only in I-frames as they contain more information then P or B frames inducing a larger capacity for similar imperceptibility and robustness such as \cite{zhoafeng2016, gaj2016, abdi2017, Kaur2019, Cai2015}. But as P and B frames use the I-frames (sometimes indirectly) to describe their frames, distortion such as the watermark applied on I-frames propagates to P and B frames damaging the quality of the video. Because of this, other schemes decided to embed only P-frames (\cite{dutta2017}). Finally, some decide to embed all frames but B-frames as they have an embedding capacity so small that the trade-off with the computational complexity is not worth it \cite{Perez-Meana10}. 

\paragraph{} The second concept is scalability as introduced by the SVC codec extension \cite{schwarz2007}: "the removal of parts of the video bit stream in order to adapt it to the various needs or preferences of end users as well as to varying terminal capabilities or network conditions". Three kinds of scalability are available with SVC:
\begin{itemize}
    \item Temporal Scalability: To adapt a data flow to the constraints, frames are spread into $n$ hierarchical levels. The transmission's frame rate is then adapted accordingly to the capabilities of the system by dropping all frames above a given layer. This way when the user's available throughput increases, he/she accepts more frames, and hence, increases his frame rate.
    \item Spatial Resolution Scalability: In the coded bitstream, multiple resolutions of the same frame are present. From the base image, copies of various resolutions are computed using decimation (\cite{test1993}). When the receiver decodes the bitstream, he/she drops all copies but the one he/she can afford to process. 
    \item Quality Scalability: This last type of scalability is slightly similar to the spatial one. As codecs all have a quantization step, quality scalability create copies of the frame with the same size processed with increasing quantization factors. 
\end{itemize}

\paragraph{} We characterize a watermark as scalability resistant if it can be detected no matter how the system adapted the transmission due to scalable compression. To achieve this, \cite{lu2013} embeds in the layer 0 of the temporal scale so that if all but one layer are dropped, as it will be layer 0, the watermark will still be there. For spatial scalability, all layers need to be able to detect the watermark. Finally, for quality scalability, the watermark simply needs to be resistant to quantization in order to be embedded in each copy.

\begin{figure}
    \begin{center}
        \includegraphics[height=600px]{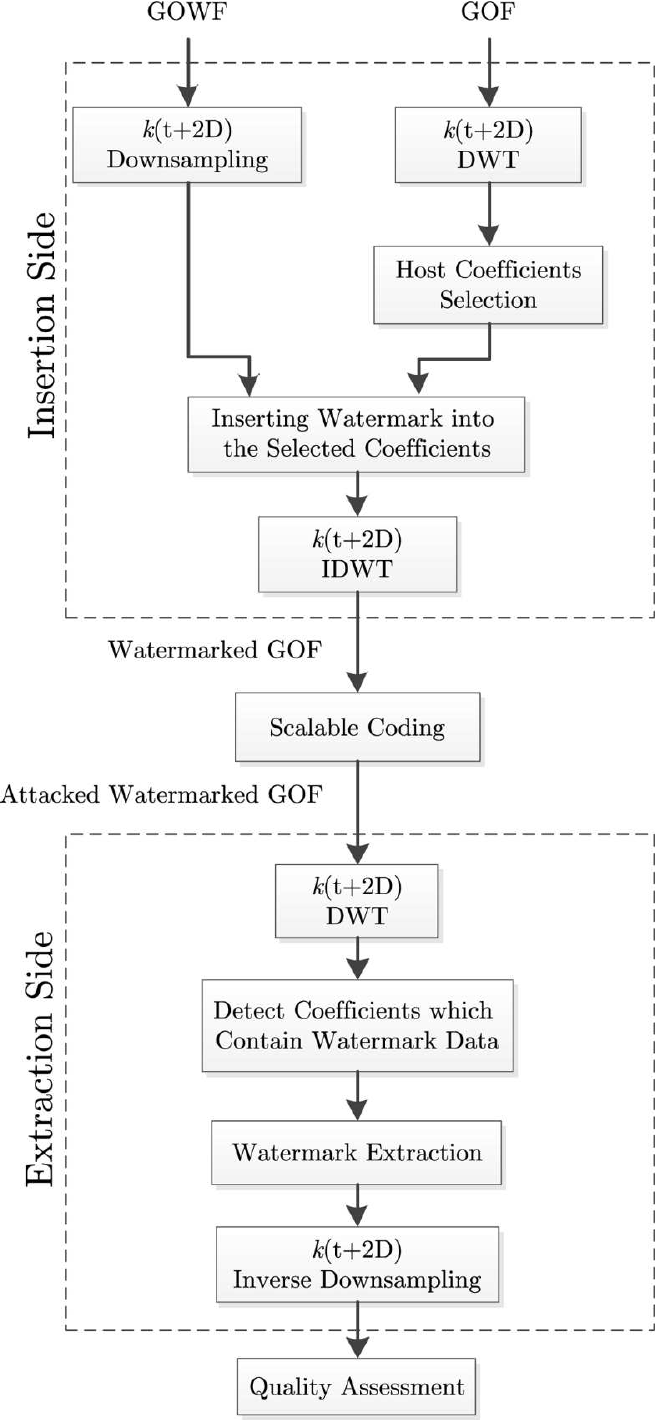}
        \caption{Scheme for scalable watermarking defined in \cite{amiri2019}}
    \end{center}
  \label{fig:scalable}
\end{figure}

\paragraph{} Some other schemes define themselves as "scalable watermark". The idea is to embed the watermark in such manner that the quality of the retrieved watermark also depends on the system's capabilities. In this scope, a technique used in \cite{abhayaratne2013} and \cite{amiri2019} is to downsample the watermark and embed it in specific DWT coefficients at various locations as shown in the flowchart of Figure \ref{fig:scalable}.

\subsection{Video as data or Network Watermarking}
\label{network}
\begin{figure}
  \includegraphics[width=\linewidth]{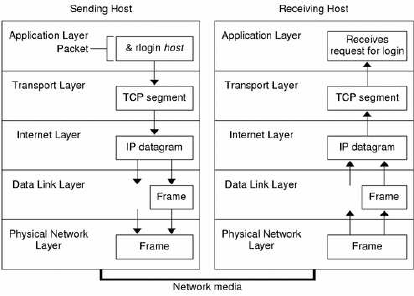}
  \caption{Layers of the TCP/IP protocol stack}
  \label{fig:tcpip}
\end{figure}

\paragraph{} We now consider that the video we are watermarking not as a stream of frames, but a stream of bits aggregated in packets of data. Those packets are then encapsulated by headers through the multiple protocol layers of TCP/IP \cite{forouzan2002tcp} (Figure \ref{fig:tcpip}). Network watermarking relies on these protocols to embed its watermark. There are two major advantages to using this kind of watermarking: 1) it can be used on any kind of media (text, video, image, audio,…) and 2) it does not affect at all the content and quality of the media.

\subsubsection{Physical Layer embedding}
\label{DSSS}
\paragraph{} At the physical layer, the data bitstream aggregated in bit frames (unrelated to image frames) is passed between nodes of the network before being transmitted as a signal. This signal can be used for embedding by slightly spreading its frequency spectrum so that this embedding actually appears as white noise since they often rely on PseudoNoise codes. Such watermarking schemes, described for example in \cite{su1998, li2013} are referred as Direct Sequence Spread Spectrum watermarking (or DSSS watermarking).

\subsubsection{Storage Channels}
\label{storage}

\begin{figure}
  \includegraphics[width=\linewidth]{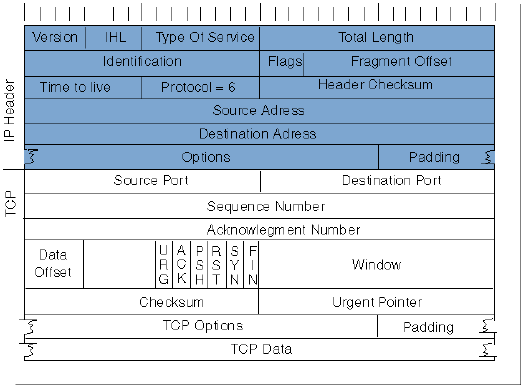}
  \caption{Headers of the TCP/IP protocols}
  \label{fig:hdr}
\end{figure}

\paragraph{} TCP and IP headers are composed of multiple fields as presented in Figure \ref{fig:hdr}. Those fields contain some redundancies that can be directly exploited in order to embed a watermark. This kind of network watermarking is known as storage channel-based watermarking, as they exploit unused fields to store the watermark information. In \cite{kundur2003} for example, the authors manage to store one bit per datagram by modifying the 3-bits flag field of the IP header. In \cite{zander2006}, the modified field is the Time To Live, using the fact that two IP datagrams from the same origin going to the same destination usually have about the same arriving TTL value. \cite{kumar2011} embeds the watermark in the TCP Sequence Numbers: by modifying the size of the segment, the author can decide the value by which the SN will have increased by the end of the transmission and use it to embed his watermark. A lot of other possibilities allow the embedding of extra data, \cite{collins2016} describe some other possible schemes such as checksum or packet length alterations.

\subsubsection{Timing Channels}
\label{timing}

\paragraph{} Watermarking using timing as the carrier signal is another kind of network watermarking which has the convenience of being quite difficult to detect \cite{iacovazzi2016network}. Therefore, it is a good solution if the watermarking is used for secret communication. Moreover, another extremely important advantage is that the embedder and detector do not even have to analyze the content of the headers nor the payload of the transmitted bit frames. It can for example be done by routers relaying the frames between two end-points. However, their undetectability has a cost paid as delays. Indeed, timing channel watermarking uses such delays to embed the information. Various methods allow such embedding. The one the most common in literature due to its simplicity is to alter the Inter Packet Delay (IPD) for the watermark insertion (\cite{houmansadr2009, iacovazzi2018, Mazurczyk2016}). But other techniques such as exploiting TCP segment temporal bursts as in \cite{luo2008} also exist.

\subsubsection{Application-Protocol Channels}
\label{AP}

\paragraph{} The application layer is the layer with the largest amount of possible protocols such as FTP, SMTP, DNS, HTTP, SSH, RTP, and countless others \cite{braden1989requirements}. The choice of this protocol directly depends on the use of the packet: IMAP and POP are protocols commonly used for mail applications, Tor for anonymity network, RTP for real-time communications and so on. These protocols have the possibility to add an optional header to the transmitted packet, which is the embedding location of AP channel-based schemes. 
\paragraph{} \cite{lucena2004} developed a system for the SSH protocol that generates MAC-like messages by simulating randomness through cryptography and replaces the MAC of the original message by the encrypted embedded message, he uses a secret field to notify the receiver that this packet's MAC is to be interpreted as a message so that it will be able to extract the hidden data, recompute the original MAC, and restore it in the SSH header. In \cite{dimitrova2017}, the author embeds the message in the number of consecutive spaces included in the headers of HTTP/1.x packets. This method relies on the fact that HTTP/1 does not limit the size of the URI in the request. Regarding RTP, \cite{mazurczyk2008} describes various ways to embed steganographic information in RTP packets, either by using the unused fields of the header or by altering the RTP security mechanisms of encryption and authentication. Multiple locations are possible for almost any application protocol.

\section{Embedding Methods for Media Watermarking}
\label{methods}

\paragraph{} In the case in which we consider the video to be a visual object, we have previously defined which blocks are to be embedded by the watermarking process. We now discuss the various ways those blocks can store the extra information.

\subsection{Spatial Domain}
\label{spatial}

\subsubsection{Least Significant Bit embedding}
\paragraph{} Least Significant Bit is an embedding method for spatial domain watermarking quite present in the literature. The color of a given pixel can be represented by a binary number as a sequence of bits. In most cases, the value of the LSB of the pixels is considered irrelevant to the visual rendering of the image. Therefore, each pixel of the image can carry at least one bit of extra information without creating a noticeable difference, which can be really convenient to embed for example a binary image or just a binary sequence representing the embedded data. \cite{bhatt2015} shows an example of the simplest implementation of this scheme of 8-bits gray scale watermark embedding on a 8-bits gray scale covert image. An example of a more advanced scheme is \cite{venugopala2014} where a watermark is embedded in selected blocks of the Y channel of the image represented by a 8-bit integer. However, the robustness of this scheme is extremely low as the LSB are often the first modified when manipulations are applied on the image. Therefore, it is rarely used by recent implementations.

\subsubsection{Linear Mask embedding}
\paragraph{} A more robust approach is to adapt the strength of the watermark to the pixel embedded depending on a mask specifying areas where the HVS is less sensitive to changes. From the embedding location defined in the previous section, we create a mask that will define the strength of the watermark. Such method is explained in \cite{yamada2016} where the final value of the pixel $i$ of frame $f$ is \[y_i'^{(f)}=y_i^{(f)}+s_i^{(f)}m_i\] where $y_i^{(f)}$ is the original value of the pixel $i$ of frame $f$, $s_i^{(f)}$ is the strength mask of the watermark at pixel $i$ of frame $f$, and $m_i$ is the bit $i$ of the watermark pattern to embed. For detection, the author averages all frames. As the watermark is here the same in all frames, the content will disappear whereas the watermark will stand out. The mask can be constant, usually with value one like in the first method proposed in \cite{nikolaidis1998}, or computed, as in \cite{lu2006, yamuna2009}.

\subsection{Frequency Domain}
\paragraph{} As seen in Section \ref{HVS}, blocks of the frames are often transformed into the frequency domain in order to detect those in which the changes are less susceptible to affect the quality of the image, but it is also possible to embed the watermark directly in this domain. The principal advantage being a high gain of robustness \cite{gaj2016}. Many transforms are possible and most will be presented here along with the embedding strategies they offer.

\subsubsection{Discrete Cosine Transform Domain}

\begin{figure}
  \includegraphics[width=\linewidth]{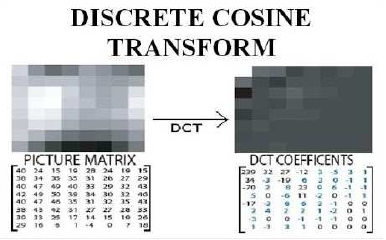}
  \caption{Discrete Cosine Transform of a 8x8 block}
  \label{fig:dct}
\end{figure}

\paragraph{} DCT is one of the most used transforms for watermarking in the frequency domain. As precised in \cite{oppenheim1999discrete}, the result of applying a DCT transform to a 8x8 block is a block of the same size composed of one Direct Current (DC) coefficient representing the average color of the region, located in the upper left corner of the block and usually much larger than the 63 other coefficients called Alternating Current (AC) coefficients representing the color changes within the block. The AC coefficients are positioned so that the low-frequency coefficients are closer to the DC coefficient and the high frequency ones further in a zig-zag manner. Because of such display, many of the down right corner coefficients have a zero-value. An example of a DCT transformed block is visible in Figure \ref{fig:dct}. 

\paragraph{} The watermark information can be embedded in any of those coefficients. As explained in \cite{chen2012}, if we embed the information in high-frequency coefficients, the changes will not affect as much the quality of the image as if we modify the low-frequency ones. However, there is a very important trade-off happening: the lower the frequency range of the embedded coefficients, the more robust the watermark will be against various attacks including noise, compression, filtering and others. Indeed, during compression, the high-frequency regions are generally modified by the quantization step (\cite{cabeen2008}). Moreover, limiting the coefficients to embed by their frequency also limit the capacity of the watermark: the less restrictive the conditions for a coefficient to be embedded are, the more slots will be available for embedding. Because of these trade-offs algorithms have been about equally developed for all AC coefficients, here are some examples:
\begin{itemize}
    \item Low-frequency: In \cite{Nakano2010}, the embedded coefficients are the $n^{th}$ lowest where $n$ depends whether the block belong to a I-frame or a P-frame. \cite{dutta2017} chooses to embed in the two lowest frequency coefficients after quantization to minimize synchronization error and the degradation of the image quality and BIR. In \cite{gaj2016RST} the five lowest frequency coefficients are used for good robustness. 
    
    \item Mid-frequency: \cite{Jadhav2014} uses 3D-DCT using time as the third dimension and embeds the watermark in some mid-range frequency coefficients whose positions represent a part of the extraction key. \cite{joshi2012} and \cite{chen2012} decide to embed the information in a continuous mid-frequency range of non-zero coefficients to balance robustness and transparency. 
    
    \item High-frequency: \cite{abdi2017} inserts the watermark in the highest frequency non-zero AC coefficients, but do so after quantization of the DCT block, this way, it will still be robust enough even in the high frequency range. In \cite{gaj2016}, the DC coefficient along with three AC coefficients at position (4;0), (0;4) and (4;4), which are usually in the high to mid-range frequency are selected for embedding.
\end{itemize}

\paragraph{} Other possibilities are for example: \cite{zhoafeng2016} that simply embeds all non-zero AC coefficients of the selected macro-blocks. Or \cite{Cai2015}, who defines a threshold value $T(\alpha)$ to describe this trade-off. This threshold can either be optimized using compressed sensing theory or chosen manually by the user specifically for his application. 

\paragraph{} The watermark implementations relying on alteration of the DC coefficients usually induce a low capacity and a poor transparency, therefore, they are rarer but still exist. \cite{xiao2008} discusses the use of such coefficients for watermarking and \cite{zeng2008,noorkami2008} implement such scheme. Moreover, \cite{raval2009, zhou2007} developed models where the watermark was inserted in both DC and some AC coefficients. Those methods always present themselves as very robust against signal processing operation such as compression, but suffer a lot from image degradation.  

\paragraph{} The last feature of the embedding left to define is how these selected coefficients are altered to store the watermark bit sequence. Three main strategies are adopted:
\begin{itemize}
    \item Magnitude threshold: To embed a '1', the coefficient is set higher than a threshold value, usually by adding a constant or variable value to the original coefficient. Similarly, to embed a '0' (or a '-1' depending on the form of the watermark signal), the coefficient is set under the threshold (\cite{gaj2016RST, zhoafeng2016, chen2012}). A special case of this strategy is to set the threshold to 0, meaning that the sign of the coefficient defines the watermark such as in \cite{zhang2007dct, xu2011}. 
    \item Magnitude parity: Another method is to modify the coefficient by rounding it up (or down) to the closest even integer to represent '0' and to the closest odd integer to embed a '1' like in \cite{raval2009}.
    \item Coefficients relationships: \cite{dutta2017, gaj2016} propose to use the relationship between two coefficients, meaning that the watermark bit is embedded in coefficients $C_0$ and $C_1$. To embed '0', $C_1$ is increased until it is higher than $C_0$ and it is decreased under $C_0$ to embed '1'. As two coefficients are needed for one watermarked bit, the capacity is smaller, but the watermarked signal usually shows less distortion and a better robustness than magnitude threshold embedding. A special case of coefficients relationship explained in \cite{joshi2015RT} uses the relationships of the five highest frequency AC coefficients with estimations of them made from the surrounding DC coefficients. Moreover, the author decides to embed one bit-plane of the full watermark by scene of the video sequence.
\end{itemize}

\paragraph{} Number of Non-Zero (NNZ) coefficients is a last well used method consisting of changing zero or close-to-zero coefficients in order to embed the watermark. This NNZ value can be used as a coefficient in order to embed one bit per block. The same three strategies than for DC or AC coefficients are possible. \cite{mansouri2010}, for example, uses NNZ coefficients relationships to watermark a signal using two blocks for one bit. The result obviously have a very low complexity, but shows a great deal of robustness.

\subsubsection{Discrete Sine Transform Domain}
\paragraph{} DST is a transform similar to DCT, with the exception of the fact that the common block size is 4x4 \cite{Kaur2019}, it does not include a DC coefficient representing the average value of the block and the other coefficients are not ordered in a specific manner. The texture analysis of DST relies on the NNZ DST coefficients. \cite{Liu2018} tries to reduce the intra-prediction drift by embedding the watermark as an error matrix $\Delta_{4\times4}$ in DST blocks in order to choose the parity of three coefficients of the block. \cite{Kaur2019} also uses another value: the number of coefficients with absolute value greater than one called ABGR1 to make the watermark robust: only blocks with ABGR1 greater than a threshold $\alpha$ are selected. The NNZ and ABGR1 can be modified using the same strategies than DCT coefficients.

\subsubsection{Discrete Wavelet Transform Domain}

\begin{figure}
  \includegraphics[width=\linewidth]{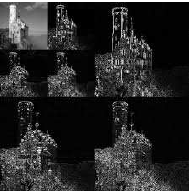}
  \caption{Level 2 Discrete Wavelet Transform of an image}
  \label{fig:dwt}
\end{figure}

\paragraph{} The DWT transform is, along with DCT, the other most used transform for watermark embedding in the frequency domain. However, techniques using DWT are often hybrids with another kind of transform. Some will be explained in Section \ref{hybrid}.

\paragraph{} DWT is a lossless transform. As explained in \cite{gupta2015}, in DWT, the matrix representing the image is decomposed in two halves. The first one represents the average coefficients of the image and the second one stores the details coefficients. For image processing, 2D-DWT is widely used. It consists of a DWT decomposing the image in two vertical halves followed by another DWT decomposing the resulting halves in two horizontal halves each. So the result is four sub-bands named after the vertical and horizontal resolution: the Low-Low (LL) sub-band in the upper left corner that represents the average of the image and the three others describe the details of the images, the High-Low (HL), the Low-High (LH) and the High-High (HH) sub-bands. Such transform can be applied multiple times called "levels" on the same image resulting in a recursive representation of the frame as shown in Figure \ref{fig:dwt} which shows an example of a level 2 DWT. As in the DCT domain, the closer of the upper left corner a sub-band is, the lower its frequency is, the more changes induce degradation to the carrier image and the more robust the resulting watermarking is.

\paragraph{} The main strategy used for embedding is magnitude threshold of selected sub-bands coefficients. In \cite{tabassum2012, barni2001}, a level 3 DWT is applied on the blue channel of the image and coefficients of high frequency sub-bands are embedded until the whole watermark is inserted. This method results in a very large capacity.The resulting robustness is relatively poor, except against frame dropping, as the watermark is fully embedded in each frame. 

\subsubsection{Discrete Fourier Transform Domain}

\begin{figure}
  \includegraphics[width=\linewidth]{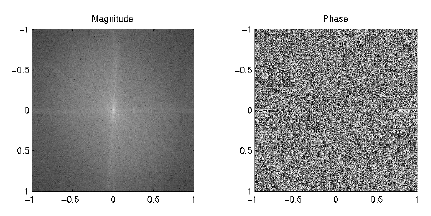}
  \caption{Representation of the magnitudes and phases of a Discrete Fourier Transform of an image}
  \label{fig:dft}
\end{figure}

\paragraph{} Applying a Fourier Transform on a matrix decompose it into DFT coefficients represented by a magnitude and a phase \cite{oppenheim1999discrete}. When processing images, unlike DCT and DST, we usually apply the DFT on the complete image instead of on a non-overlapping block decomposition of the image. Phase and magnitude of the DFT coefficients of an image are represented in Figure \ref{fig:dft}. Even though both magnitude and phase are usable, image processing usually only alter the magnitude of the coefficients. Indeed, the magnitude yields much more information about the spatial structure of the image. As in the DCT domain, a DC coefficient representing the average brightness located at the center of the block, and as we go further from the center, the frequencies of the coefficients increase. In order to improve the computational complexity of system using such transform, the Fast Fourier Transform can be used instead of the regular one.

\paragraph{} As in the previously described transforms, the higher the frequencies of the modified coefficients are, the less distortion is generated in the watermarked image. \cite{pun2006} proposes a scheme embedding a signal in the first $K$ coefficients of an image. \cite{DJUROVIC2001} skips the first $L$ coefficients and embeds the next $M$ ones to moderate the resulting distortion. \cite{urvoy2014} uses both the magnitude and the phase for insertion: the strength of the watermark embedded in the magnitude directly depends of the phase of the coefficients. 

\paragraph{} More original watermarking schemes exists using DFT or variants of DFT such as \cite{LIU2010} that combines 1-D DFT with Radon transform to embed a fence-shaped watermark into the frames with the highest temporal frequencies. \cite{Deguillaume1999} uses 3D-DFT with time as a third dimension to embed the watermark in the mid-band frequencies and obtain a compromise between visibility and robustness to lossy compression. Finally, \cite{wang2013} develops a watermarking scheme in the quaternion Fourier transform domain that uses complex numbers theory for embedding.

\subsubsection{Singular Value Decomposition}

\paragraph{} If SVD is not actually in the frequency domain, it allows to represent the image in a non spatial manner and some watermarking schemes exploit this particularity, hence its presence in this section.

\paragraph{} In SVD for image processing \cite{sadek2012svd}, a square image $I$ represented as a $n\times n$ matrix is decomposed as $I=USV^H$ where $S$ is a diagonal matrix containing the singular values of $I$ that can be modified for watermark embedding. For example, \cite{liu2002} describes a scheme where the matrix $S+\alpha W$ (where W is the watermark matrix and $\alpha$ the strength of the watermark) is decomposed into $U_WS_WV_W^H$ and the watermarked image $A_W$ is obtained with $A_W=US_WV^H$. An interesting property of such embedding is its non-inversibility that we mentioned in the definition of ambiguity attacks. The watermark can however also be embedded in the U and V matrix of the decomposition as presented in \cite{CHUNG2007}. An important counter effect of SVD embedding is that the watermark signal is usually retrieved with a relatively high distortion.

\subsubsection{Hybrid Domains} 
\label{hybrid}
\paragraph{} More recently, many schemes combined previously described techniques to implement hybrid domain watermarking schemes. Most of those schemes use DWT, SVD and/or DCT. However, for most of those schemes, the computational complexity is drastically increased as these transforms are computationally heavy and multiple transformations are used in those methods.

\paragraph{} In DWT-SVD, a DWT is applied to the image, then a SVD embedding is executed on some of the resulting sub-bands. When \cite{Kaur2015} decides to embed only the HH sub-band for better transparency, \cite{ganic2004} embeds all four sub-bands to achieve better robustness and watermark fidelity.

\paragraph{} For DWT-DCT embedding, some schemes such as \cite{hazim2019} use a DCT embedding in the mid-range frequency coefficients of one or multiple DWT sub-bands. Some others like \cite{Abdulrahman2019} apply a DWT embedding in all four sub-bands of a DCT transformed image. In another kind of DWT-DCT watermarking described in \cite{sharma2017, Jiansheng2009}, the DCT coefficients of the watermark are embedded into the DWT coefficients of the carrier signal in an attempt to increase the robustness of the algorithm.

\paragraph{} Finally, some schemes such as \cite{Zear2018} even combine the three transforms: first a level 3 DWT is applied, then a DCT on the LH1, then a SVD embedding on the $S$ component of the resulting coefficients.

\subsubsection{Other Domains}

\paragraph{} The number of transforms that can be used for image processing in very important. Therefore, they can not all be described, but we can however mention some others:
\begin{itemize}
    \item the Slantlet Transform: It is considered as an extension of the DWT and similarly generates four frequency sub-bands. \cite{Alwan2011} is an example of watermarking is this domain where mid-range frequencies are embedded using the magnitude threshold strategy. Also \cite{mohammed2012} embeds its watermark in the HL and LH sub-bands by altering coefficients relationships. The two coefficients used are the values of the same pixel in the mentioned sub-bands.
    \item The Shearlet Transform: Its particularity is that it describes an image using multiple directions of its singularities. Such transform is explained in more details and used for watermarking in \cite{ahmaderaghi2018} combined with statistical decision theory.
    \item The Contourlet Transform: Based on similar principles as the Shearlet transform, it is used by \cite{li2019} to embed a watermark in the low-frequency sub-band in order to be robust against geometric attacks.
\end{itemize}

\subsubsection{Motion domain}
\paragraph{} In the case of a video, as previously said, the motion vector can be used to select the embedding blocks, however, this motion vector can also be used as embedding material: \cite{zhang2001} uses the phase angle of the motion vector to insert the watermark information.

\section{Watermarking and other technologies} 
\label{technologies}
\paragraph{} Some technologies have been combined with digital watermarking in attempts to enhance the security, the robustness, the imperceptibility and to adapt watermarking to new fields, like to new types of objects such as 3D videos (\cite{Waleed2015}) or holographic images (\cite{li2019holo}) that will not be detailed here.   

\paragraph{Homomorphic Encryption}
Homomorphic encryption allows to modify an encrypted document without the need to decrypt it \cite{vaudenay2004communication}, and hence, knowing its content. The use of such technology along with watermarking has many extremely interesting applications: one could use a third party to watermark their media without ever having to reveal to them the original media, or add watermarks to an end-to-end encryption system. This combination has been studied in \cite{ABDALLAH2016} with watermarking with a Singular Value Decomposition method, or in the DWT domain as in \cite{guo2015}. However, the computational complexity of such method it to take into account as this technology is still relatively young.

\paragraph{Machine Learning}
Along with Deep Learning, ML is extremely popular in the current data science researches. The quality of the classifiers that can be achieved by such technologies using Support-Vector Machine or/and Least-Square can find interesting applications for watermark detectors in statistical embedding schemes as exposed by \cite{wang2013}. Another use of ML is to optimize the parameters of a scheme in order to find the best compromise between time complexity, robustness and transparency as in \cite{ABDELHAKIM2018}. Related to its use for detection, such technology can however also be used to break watermarks as in an oracle attack such as exposed in \cite{quiring2018ML} to remove watermarks using adversarial learning.

\paragraph{Fractal Coding}
Used in some compression methods, fractal coding is based on the repetition of objects in an image to reduce its weight \cite{monro1992fractal}. \cite{channapragada2016} proposes a procedure to embed two bit-planes of the blue component of an image that are coded using fractal coding theory and embedded with a watermark. This scheme was designed by studying previous implementations proposed by  \cite{KIANI2011, Candik2000}. 

\paragraph{Quantum Watermarking}
Without going into details due to its complexity, many researches also aim at designing watermarking schemes for quantum signal such as \cite{Heidari2016} that embeds a quantum image using the LSB method, \cite{Zhang2013} that decided to use Quantum Fourier Transform to preserve the carrier image's visual or \cite{Yang2014} that uses the Quantum Wavelet Transform instead.

\paragraph{Blockchains}
As another popular field of computer science, the pairing of blockchains with watermarking has also been studied: to be able to retrace the transaction trails and modifications history of a media through its life as in \cite{bhowmik2017}. To secure the watermarked information or confirm the watermarks creation order for multiple copyrights management (\cite{meng2018, zhaofeng2018}). Or finally as part of a large secure authentication scheme as defined in \cite{liu2020} for medical use of cloud-based image management.

\section{Conclusion}
\paragraph{}
Digital Video Watermarking is security mechanism used in a wide range of applications including copyright protection, tampering identification, clandestine communication, traffic analysis and access control. This mechanism can be evaluated regarding two main criteria: visibility and robustness. In order to balance those characteristics to satisfy the application's needs, the developer can select a precise location for the embedding of the watermark. Indeed, he can decide to make it as little detectable as possible by either selecting an area using the definition of the Human Visual System or by choosing to embed the watermark in the information surrounding the video. YouSkyde (\cite{Mazurczyk2016}) uses for example package dropping to implement a hidden channel in a Skype communication whereas \cite{CEDILLOHERNANDEZ201440} insert some information in a selected area of the video frames chosen based on four HVS criteria. In the case where visual embedding is chosen, the watermarking scheme can be categorized depending on the embedding process used. The first category is insertion in the spatial domain that include LSB modification and linear masking as \cite{lancini2002robust}, that uses spatial watermarking for copyright protection and embedding of indexing information. It has the advantage of being relatively easy to implement and quite fast as only few operations are required for the embedding. The second category is insertion in the frequency domain, mainly the DCT, DST, DWT, DFT and SVD domains but also in combination of those. This method usually allows to achieve better robustness and invisibility as the watermark information can be better spread on the irrelevant coefficients of the image. For example, Hazim et al. (\cite{hazim2019}) uses a hybrid transform of DWT and DCT, as the frequency domain allows a increased capacity and those two domains lead to a good robustness for the watermark which could be extremely relevant if the signal can be heavily degraded such as with poor network conditions. 
\paragraph{}
While there is a lot of published results about pairing watermarking with innovative and emergent technologies, it feels like the application of watermarking for Real Time Communication has been so far overlooked. Most of the research we could find that take speed into account usually limit themselves to evaluate the time performance of detecting or extracting the watermark. For real-time communication, where the entire time budget for all processing applied to a frame (encoding, watermarking, encryption, packetization, transport, ...) is limited by the acquisition speed (33ms at 30fps, 16ms at 60fps) the embedding speed is even more important than the detecting speed. As far as we know, Robust watermark scheme allowing a watermark embedding in a such short time is still to be developed.  It could have interesting uses such as securing video chat, screen sharing or other streaming and broadcasting applications.

\bibliographystyle{unsrt}  
\bibliography{references} 
\end{document}